\documentclass[12pt]{iopart}
\usepackage[dvips]{graphicx}

\begin{document}

\title[Two dimensional self-avoiding walk with hydrogen-like
bonding]{Two dimensional self-avoiding walk with hydrogen-like
bonding: 
Phase diagram and critical behaviour}

\author{D P Foster\dag and F Seno\ddag}
\address{\dag Laboratoire de Physique Th\'eorique et 
Mod\'elisation (CNRS ESA 8089),
Universit\'e de Cergy-Pontoise,
5~Mail Gay-Lussac 95035 Cergy-Pontoise Cedex, France\\
\ddag INFM--Dipartimento di Fisica, Universit\`a di Padova, Via
Marzolo 8, 35131 Padova, Italy}

\begin{abstract}
The phase diagram for a two-dimensional self-avoiding walk model on
the square lattice
incorporating attractive short-ranged 
interactions between parallel sections of
walk is derived using numerical transfer matrix techniques. The model
displays a collapse transition. In contrast to the standard
$\theta$-point model, the transition is first order. The phase diagram
in the
full fugacity-temperature plane displays an additional 
transition line,
when compared to the $\theta$-point model, as well as a critical
transition at finite temperature in 
the hamiltonian walk limit.
\end{abstract}



\maketitle

\section{Introduction}

Self-avoiding walks have been widely studied as models of polymers in
dilute solution\cite{genbks}. 
A self-avoiding walk  on a lattice is defined as a random walk
 which is forbidden from visiting the same lattice-site more
than once\cite{genbks}. 
In the limit of very long walks it models a polymer in a
good solvent. In order to model the relative affinity between
the monomers (compared with the solvent) an attractive energy is
introduced
between non-consecutively visited nearest-neighbour sites\cite{cont,cont1}. 
This is the
standard $\theta$-point model. 
At high temperatures the polymer is 
happy to be in solution.
As the temperature is lowered the
polymer tends to collapse in on itself, and at a given temperature
will precipitate from solution. 
For an idealised infinitely long
polymer the high temperature and low temperature regimes are separated
by a phase transition occurring at a temperature known as the theta
temperature, $T_{\theta}$\cite{genbks,flory,degennes1}. 
In the grand-canonical description 
 the average
length of the walk is controlled by a fugacity $K$ (or chemical
potential $\mu$ where $K=\exp(-\beta\mu)$). 
At high temperatures the (average)
length of the walk diverges smoothly as a critical fugacity $K_c(T)$ is
approached. At low temperatures the length jumps discontinuously at a
given value of the fugacity $K^*(T)$ (corresponding to a first order
transition). The two lines $K_c(T)$ and $K^*(T)$ follow continuously
one from the other, and the two behaviours are separated by a
tricritical point, at a temperature corresponding to $T_{\theta}$. The
transition line $K_c(T)$ then $K^*(T)$ is identified with the
``thermodynamic limit'' of the polymer problem. Along this line, the
proportion of visited sites is zero for $T<T_{\theta}$ and becomes
non-zero continuously for $T>T_{\theta}$.  

While it is clear that the main experimental interest lies in three
dimensions (although there are some experimental results also in two
dimensions\cite{expt2d}), there has been  
a lot of theoretical interest in the two 
dimensional case\cite{ds,ssv1,merio}. 
This is in part due to the fact that the upper
critical dimension for a tricritical point is three\cite{lawrie}. 
In dimensions
lower than three (but larger than one) there is a possibility for
a greater diversity of critical behaviours.
 
\begin{figure}
\begin{center}
\includegraphics[width=7cm,angle=-90]{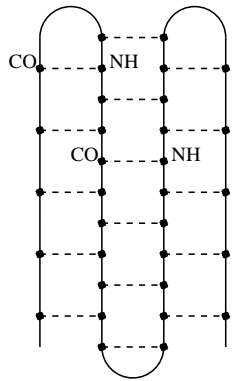}
\end{center}
\caption{A schematic representation of hydrogen bonds in a beta sheet
(following Bascle \etal\cite{bascle})}\label{betasheet}
\end{figure}

In this article we investigate a two-dimensional 
model with a restricted set of interactions, compared to the
theta model; an interaction is only present between
nearest-neighbour non-consecutively visited sites belonging to	 
straight portions of polymer, i.e. an interacting site cannot sit on
a corner of the walk (see \Fref{interact}). 
This choice of interactions may be motivated by
the r\^ole of hydrogen bonding in the formation of secondary
structures in proteins\cite{bascle}. 
These bonds are formed between the \textbf{CO}
group of one peptide and the \textbf{NH} group of a nearby
(non-consecutive) peptide. This bond imposes a constraint on the 
orientation of the relevent peptides. This type of bonding 
is thought to be responsible for the
formation of alpha helices and beta sheets\cite{pauling}. In our
two-dimensional toy-model
the formation of alpha helices is not possible.
A representation of the formation of a beta sheet through hydrogen
bonding is shown in \Fref{betasheet}.

Our main interest in this paper is not to give a realistic model for protein
folding, but rather to understand the influence of these apparently small
modifications to the local attractive interactions on the critical
behaviour of lattice interacting self-avoiding walks.
To this end we  present the phase diagram and elucidate the
critical behaviour in the full fugacity---temperature plane.

In the following section we present in detail the model as well as
the transfer matrix calculation of the
quantities of interest. In section~\ref{PD} we present the phase
diagram and discuss the different transitions present. In
section~\ref{hamilt} we concentrate more specifically on the
Hamiltonian Walk limit of the model in which 
all sites are visited exactly
once. In section~\ref{CONC} we finish with some concluding
remarks.

\section{The model and the transfer matrix calculation}\label{model}

\begin{figure}
\begin{center}
\includegraphics[width=7cm]{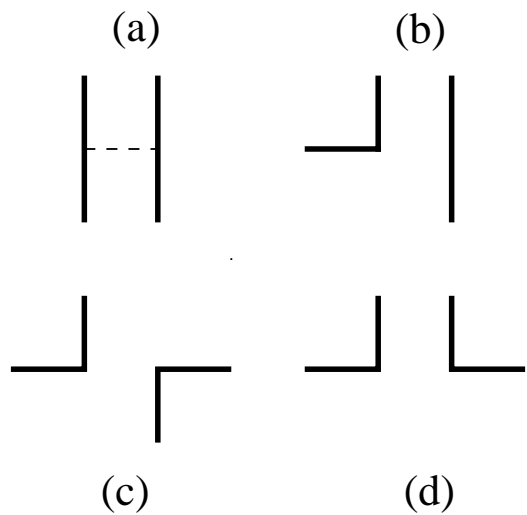}
\end{center}
\caption{Possible configurations of nearest-neighbour non-consecutively
visited sites. Configurations of type (a) include an attractive
interaction energy $-\epsilon$, configurations of types (b), (c) and
(d) interact in the standard $\theta$-point model, but here do not
participate in hydrogen type bonding.}\label{interact}
\end{figure}

The model studied here consists of a self-avoiding walk embedded on a
square lattice. An interaction energy $-\epsilon$ is assigned for each
non-consecutively visited pair nearest-neighbour sites for which the
four bonds are parallel (see \fref{interact}). A chemical potential,
$\mu$, and related  fugacity
$K=\exp(-\beta\mu)$ are associated to each step.

The grand canonical partition function is then 
\begin{equation}\label{partfn}
{\cal Z}=\sum_{{\rm walks}} K^{N} \exp(\beta N_I \epsilon),
\end{equation}
where $N$ is the number of steps in the walk and $N_I$ is the number of
interactions. 
Quantities of interest may be calculated as appropriate
derivatives of the partition function, for example, the density is
given by:
\begin{equation}\label{length}
\rho=\frac{\langle N \rangle}{\Omega}=\frac{K}{\Omega}\frac{\partial \log{\cal 
Z}}{\partial K},
\end{equation}
where $\Omega$ is the number of lattice sites.
The relative length fluctuations are related to the 
derivative of $\rho$ through:
\begin{equation}\label{lfluct}
\frac{\langle N^2 \rangle - \langle N \rangle^2}{\langle N \rangle} =
\frac{K}{\rho}\frac{\partial \rho}{\partial K}.
\end{equation}.

In the dilute polymer regime (small enough $K$) 
the
correlation function 
may be identified with the probability that two given points are
joined by a  walk. 
In this  case the correlation length must scale in
the same way as any reasonable characteristic length scale associated
to the walk. It is usual to take the radius of gyration, which
corresponds to the average squared distance the walk steps from the
common centre of gravity. It follows that 
\begin{equation}
R_G\sim |K-K_c|^{-\nu},
\end{equation}
where $\nu$ is defined as the standard correlation length
exponent\cite{stanley} .
Using the relation \eref{length}, one finds 
\begin{equation}\label{rgnu}
R_G\sim\langle N \rangle^{\nu},
\end{equation}
from which we may see that the walk is fractal; the mass of the walk
scales linearly with $N$, and hence $R_G^{1/\nu}$. The fractal
 (Haussdorf)\cite{Man}
dimension of
the walk is then $d_H=1/\nu$\cite{genbks}. 
It is usual in such problems to take
equation \eref{rgnu} as the definition of the exponent $\nu$\cite{genbks}. 
In this
case, note that it only relates to the usual critical
exponent in the case of a critical transition in the dilute regime. In
the collapsed phase(s) the polymer fills the lattice, the mass scales
as $R_G^d$ and equation \eref{rgnu} gives $\nu=1/d$.

In this article we propose to use a transfer matrix formalism to
calculate ${\cal Z}$ as a function of $K$ and
$\beta\epsilon$\cite{tmat,derh,ders}. 
For convenience we
shall
 from now on set
$\epsilon=1$, which simply corresponds to a choice of temperature
scale. 

The idea behind the transfer matrix formalism is to calculate the
partition function on a lattice which is infinite in one direction
(the  $x$ direction, say) but finite in the other ($y$) direction. The
thermodynamic limit is attained by increasing the lattice width. 

The standard way of considering the problem is as follows: Define the 
restricted 
partition function $Z_x({\cal C}_0,{\cal C}_x)$ as the
partition function for a portion of walk between the origin and
$x$. The walk has a column state ${\cal C}_0$ at the origin and ${\cal
C}_x$ in column $x$. One may then write the following recursion
relation:
\begin{equation}\label{recpart}
Z_{x+1}({\cal C}_0,{\cal C}_{x+1})=\sum_{{\cal C}_x} 
Z_x({\cal C}_0,{\cal C}_x){\cal T}({\cal C}_x,{\cal C}_{x+1}),
\end{equation}
where ${\cal T}({\cal C}_x,{\cal C}_{x+1})$ is the additional 
Boltzmann weight to add column $x+1$ in configuration ${\cal C}_{x+1}$
next to column $x$ in configuration ${\cal C}_x$. This forms a
(transfer) matrix.

\begin{figure}
\begin{center}
\includegraphics[width=7cm]{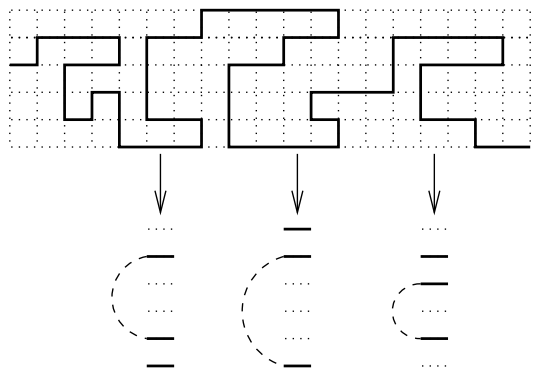}
\end{center}
\caption{Example of a polymer configuration with the corresponding
column states}\label{colconf}
\end{figure}

That this recursion should be valid for a spin system is fairly
straightforward, since the interactions are all local. For a polymer
model it is less clear that this should be possible, since one has to
take into account non-local factors, most notably one has to ensure
that the partition function describes only one chain, without the
formation of ``orphan'' loops. This is done by appropriately defining
the column states ${\cal C}_x$. For the self-avoiding walk problem
with no added interactions (i.e. $\epsilon=0$) it is
sufficient\cite{tmat} to define a column configuration by the
arrangement of horizontal bonds in the column along with information
about the connectivities between the bonds, that is information about
which pairs of horizontal bonds are connected by polymer loops to the
left (taking $x$ as increasing towards the right). See \fref{colconf}
for an example.  By successive application of \eref{recpart}, one
finds
\begin{equation}
Z_x({\cal C}_0,{\cal C}_x)=\langle {\cal C}_0|{\cal T}^x|{\cal
C}_x\rangle.
\end{equation}

The situation becomes a little more complicated if interactions are to
be taken into account. 
It is
necessary to know the column states over three columns in 
order to incorporate the horizontal bonds. The
appropriate modification to the above equations is\cite{derh,yeomv}:
\begin{equation}
Z_{x+1}({\cal C}_0,{\cal C}_1;{\cal C}_{x},{\cal C}_{x+1})
=\sum_{{\cal C^{\prime}}_{x-1},{\cal C^\prime}_x} 
Z_x({\cal C}_0,{\cal C}_1;{\cal C^{\prime}}_{x-1}{\cal
C^{\prime}}_{x}){\cal T}({\cal C^{\prime}}_{x-1},{\cal C^{\prime}}_{x};
{\cal C}_x,{\cal C}_{x+1}),
\end{equation}
and
\begin{equation}\label{partf}
Z_x({\cal C}_0,{\cal C}_1;{\cal C}_{x-1}
,{\cal C}_x)
=\langle {\cal C}_0,{\cal C}_1|{\cal T}^{x-1}|{\cal C}_{x-1},{\cal
C}_x\rangle.
\end{equation}
The primed configurations are introduced to retain the matrix
notation, the elements of the matrix are now labelled by the
two column states for the input and two for the output (primed),
with the restriction that only matrix elements with ${\cal C}_x={\cal
C^\prime}_x$ may be non-zero.

It is required that the partition function sum over walks of all
possible lengths. To do this we define a  
restricted partition function ${\cal Z}_x$ for all the walks
having a caliper extension $x$, i.e. walks which are
entirely 
confined to a strip of length $x$, and extend over the entire length
of the strip. This is simply
$Z_x$ summed over all configurations ${\cal C}_0$, ${\cal C}_1$, ${\cal
C}_{x-1}$ and ${\cal C}_x$ compatible with the starting and ending
configurations of the walk. The full partition function is then given
by
\begin{eqnarray}
{\cal Z}&=&\sum_{x=0}^{\infty} 
{\cal Z}_{x+1},\\
&=& \sum_{b.c.}\,^{\prime} 
\sum_{x=0}^{\infty} {\cal T}^x
\end{eqnarray}
where the  primed sum is over the boundary configurations, as described
above.

Let us denote by $\{\lambda_i\}$ the eigenvalues of ${\cal T}$,
numbered in order of their moduli.
The  eigenvalue of largest modulus, $\lambda_1$ ,
is always positive. 

For small enough $K$,
$\lambda_1\leq 1$. In this case the partition function
may be written 
\begin{equation}
{\cal Z}=\sum_i \frac{\alpha_i}{1-\lambda_i(K,\beta\epsilon)}.
\end{equation}
The $\alpha_i$ are essentially constants
dependent on the boundary conditions,
which gives for the density:
\begin{equation}
\rho=\frac{K}{\Omega}\sum_i\frac{1}{1-\lambda_i}\frac{\partial
\lambda_i}{\partial K}.
\end{equation} 
If the derivative of $\lambda_i$ is finite, which is the case at least
for small enough $\beta$, the density is zero as long as
$\lambda_1<1$ and may only change from its zero value when the average
length of the walk diverges, i.e. $\lambda_1\geq 1$\cite{ders,yeomv}. 
The point
$\lambda_1=1$ is identified with the critical transition, and is often
used
in polymer science to define the ``thermodynamic limit'' of ``very
long'' polymers\cite{genbks}. 
Another way of reaching to the same conclusion is to consider two
points as being correlated if they 
are joined by a walk. One may then
identify $Z_x$ with a 
two point correlation function, and hence 
identify the correlation length, $\xi$\cite{ders}: 
\begin{equation}\label{corl}
\xi=\frac{1}{\log\left(\frac{1}{\lambda_1}\right)}.
\end{equation}
The correlation length diverges as $\lambda_1 \to 1$, consistent with
the identification of a critical transition.

The above argument needs to be treated with caution. 
As long as the lattice width is finite, the system is equivalent
to a one-dimensional system with a
finite, if large, number of states per site (each site corresponding
to a column in our model). 
It is well known
that a one dimensional system with a finite number of states per site
cannot  have an equilibrium phase
transition\cite{pierls}.
 
The reason for this apparent discrepancy is that the definition of
$\xi$ \eref{corl}
is not valid in the large $K$ phase (dense polymer phase). 
In a spin model the partition function would have been given by
${\cal Z}_x$ in the limit $x\to\infty$. The
transfer matrix would contain all the possible configurations, and a
standard analysis shows that\cite{jmy}:
\begin{equation}\label{normxi}
\xi=\frac{1}{\log\left(\frac{\lambda_0}{|\lambda_1|}\right)},
\end{equation}
where $\lambda_0$ is the largest eigenvalue, and $\lambda_1$ is the
second largest (in modulus). 
To ensure the correct construction of a
single walk all configurations consisting of an empty column
have been excluded from the transfer matrix, but replaced by the sum
over all possible polymer extensions $x$. The consequence 
of this choice
is that the largest eigenvalue in the zero-density phase has been
eliminated. Its value, as can be seen by inspection of the expression
for $\xi$, would be $\lambda_0=1$.

This prescription only holds as long as $\lambda_1<1$. If
$\lambda_1>1$ then the partition function, density and $\xi$ 
are given by:
\begin{eqnarray}
{\cal Z}
&\sim&\lim_{x\to\infty} \lambda_1^x,\\\label{normrho}
\rho&=&\frac{K}{L\lambda_1}\frac{\partial \lambda_1}{\partial K}\\
\xi&=&\frac{1}{\log\left(\frac{\lambda_1}{|\lambda_2|}\right)},
\end{eqnarray}
where $L$ is the lattice width.

For convenience we impose the condition that the walk of extension
$x^\prime$ in the $x$ direction
originates at $x=0$ and terminates either at $x=0$ or
$x=x^\prime$. As is usual in equilibrium phase transitions, the
precise choice of boundary condition does not change the
critical behaviour. Imposing this restriction it is easy to see that
column configurations with even numbers of horizontal bonds may only
follow column configurations with even numbers of horizontal bonds,
and likewise for configurations with an odd number of horizontal
bonds. This block diagonalisation is convenient since it transpires
that for much of the phase diagram 
$\lambda_1$ and $\lambda_2$ arise in different blocks. It is
relatively easy to calculate the largest eigenvalues of a matrix using
the power method\cite{power}.

\section{The $K-\beta$ phase diagram}\label{PD}

The critical lines may be identified using
phenomenological renormalisation\cite{mpn76}. At a critical
point the correlation lengths for two strip widths, measured as a
fraction of the strip width, must be the same. This reflects the scale
invariance of a critical system. In practice a finite width system is
always off-critical, however
\begin{equation}\label{fss} 
\frac{\xi_L(K^*)}{L}=\frac{\xi_{L^{\prime}}(K^*)}{L^{\prime}}
\end{equation}
gives a finite-size estimate of the critical fugacity, $K^*$, which will
tend to the true critical fugacity as the strip widths tend to
infinity.
Assuming that the system is sufficiently close to the true critical
point, the correlation length behaves (to leading order) as
\begin{equation}\label{corsc}
\xi_L=A|K^*-K_c|^{-\nu}.
\end{equation}
Since it is the finite width of the system which prevents the
correlation length from diverging, at $K^*$ we have 
$L\propto\xi_L$. If these
two scaling laws are admitted, then a finite-size estimate of the
correlation length exponent may be calculated from the equation:
\begin{equation}\label{nu}
\frac{1}{\nu_{L,L^{\prime}}}=\frac{\log\left(\frac{{\rm d}\xi_L}{{\rm
d} K}/\frac{{\rm d}\xi_{L^{\prime}}}{{\rm
d} K}\right)}{\log\left(\frac{L}{L^{\prime}}\right)}-1.
\end{equation}

\begin{figure}
\begin{center}
\includegraphics[width=7.5cm,angle=-90]{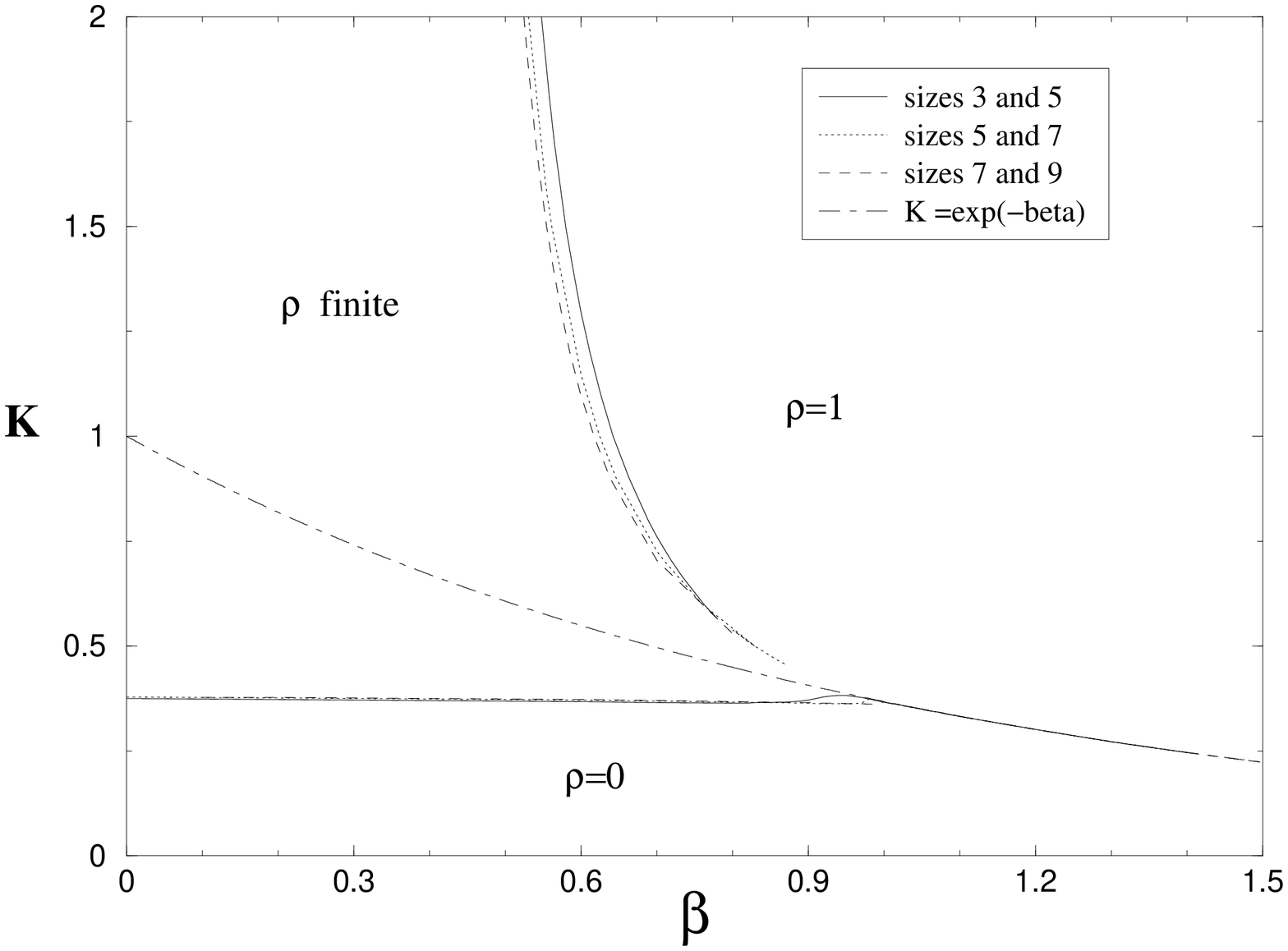}\\
\includegraphics[width=7.5cm,angle=-90]{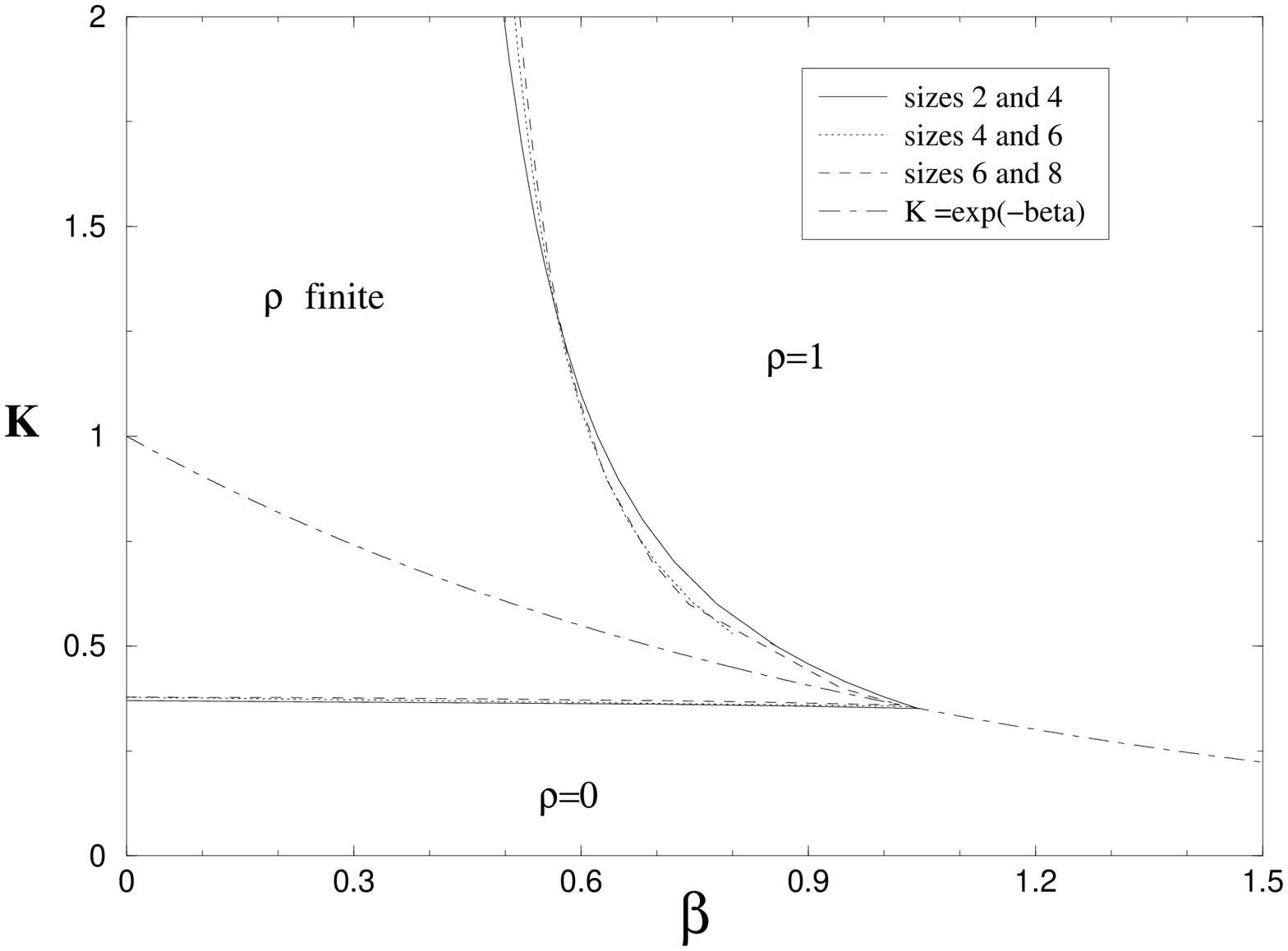}
\end{center}
\caption{Phase diagram estimates found using phenomenological
renormalisation group for odd lattice widths (top) and even lattice
widths (bottom)} \label{prgpd}
\end{figure}

There exists at least one phase transition. In the case $\beta=0$ this
corresponds to the standard Self-Avoiding Walk transition. 
As $\beta$
is increased, this point extends into a line of critical points. In
analogy to the standard $\theta$-point model, we would expect this
transition to change to first order at some given value of
$\beta=\beta_H$. As discussed in the previous section, starting from
the low-K phase, the correlation function is given by \eref{corl}. In
the high-K phases the appropriate form of $\xi$ is given by
\eref{normxi}. The phase diagram estimates found using phenomenological
renormalisation, using the appropriate form of $\xi$, are shown in
\fref{prgpd}. As is usual in this sort of problem, there are strong
parity effects, notably for the odd lattice widths $\lambda_1$ is
in the odd sector of the transfer matrix,
while for the even lattice widths $\lambda_1$
is taken from the even sector. The next largest eigenvalue,
$\lambda_2$, is in the other sector for the low-$K$ phase and for the
high temperature high-$K$ phase, while it is in the same sector for
the high-$K$ low temperature phase. This observation was not used to
estimate the transition lines, which were however found assuming that
$\lambda_1$ and $\lambda_2$ came from different sectors of the
transfer matrix. 
Not all the transition lines are found
for even lattice widths. 

\begin{figure}
\begin{center}
\includegraphics[width=7.5cm,angle=-90]{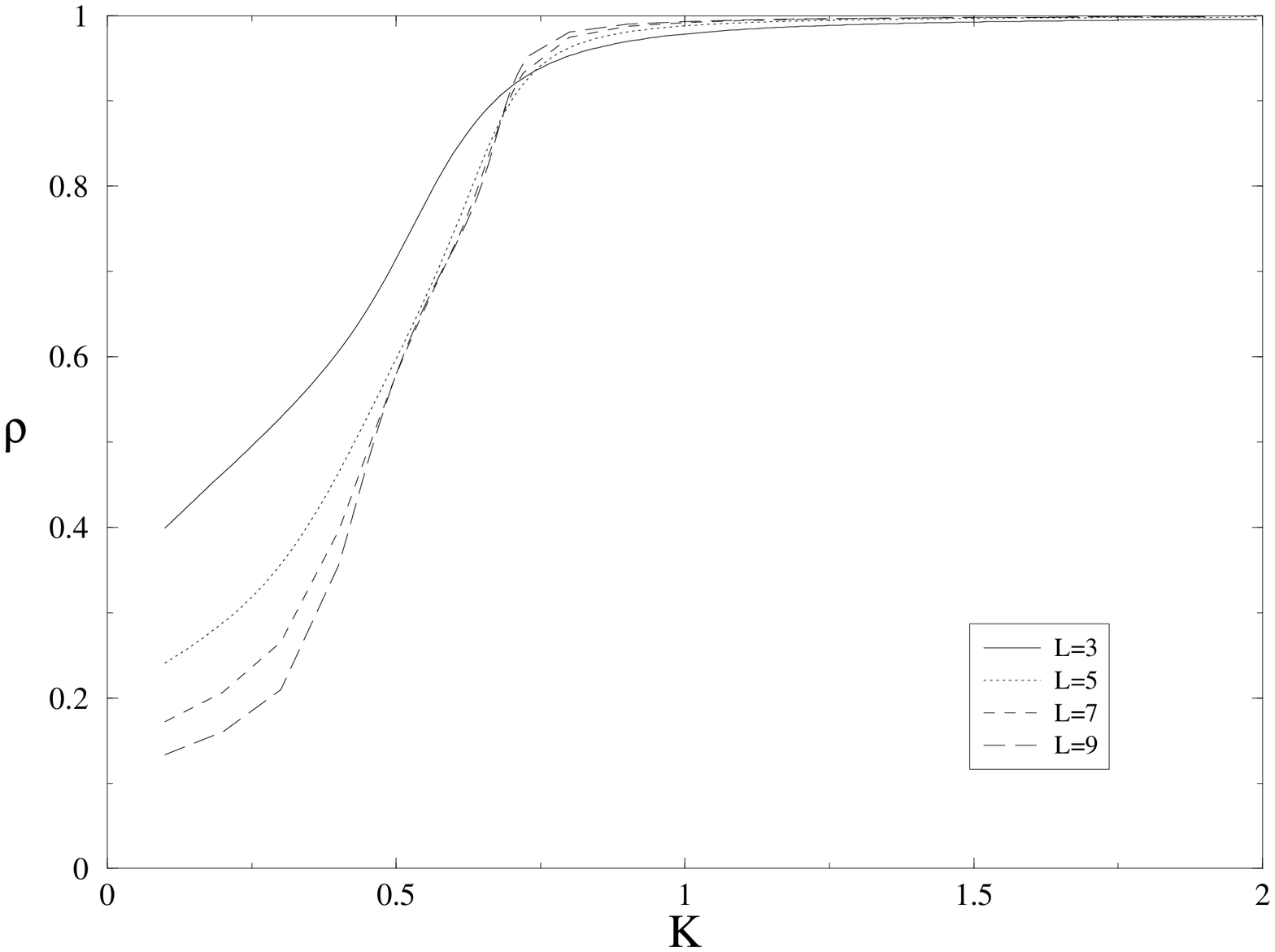}\\
\includegraphics[width=12.8cm]{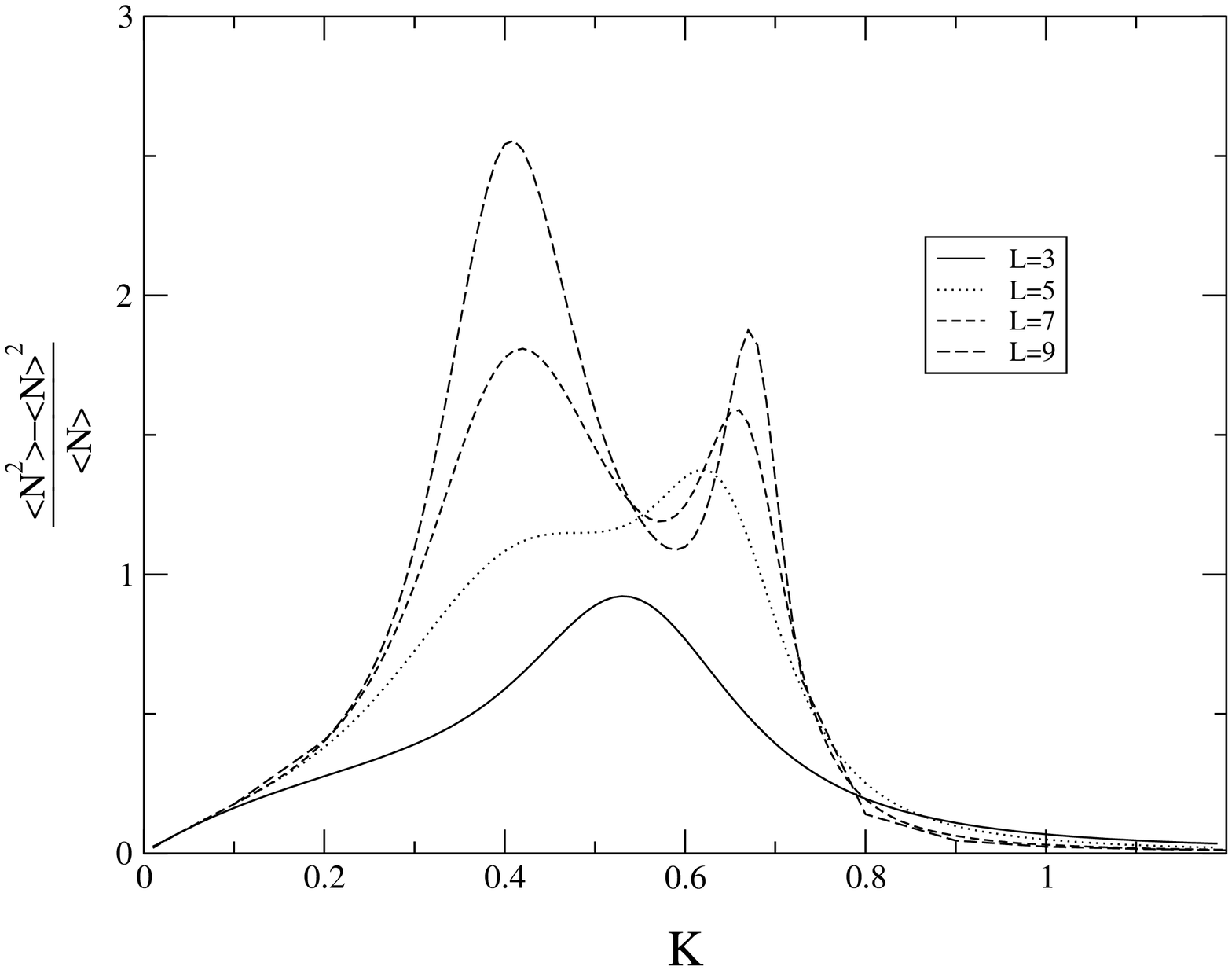}
\end{center}
\caption{Plots of density, $\rho$ (top), and density fluctuations
(bottom) for
the inverse temperature $\beta=0.7$ }\label{denspl1a}
\end{figure}

\begin{figure}
\begin{center}
\includegraphics[width=7.5cm,angle=-90]{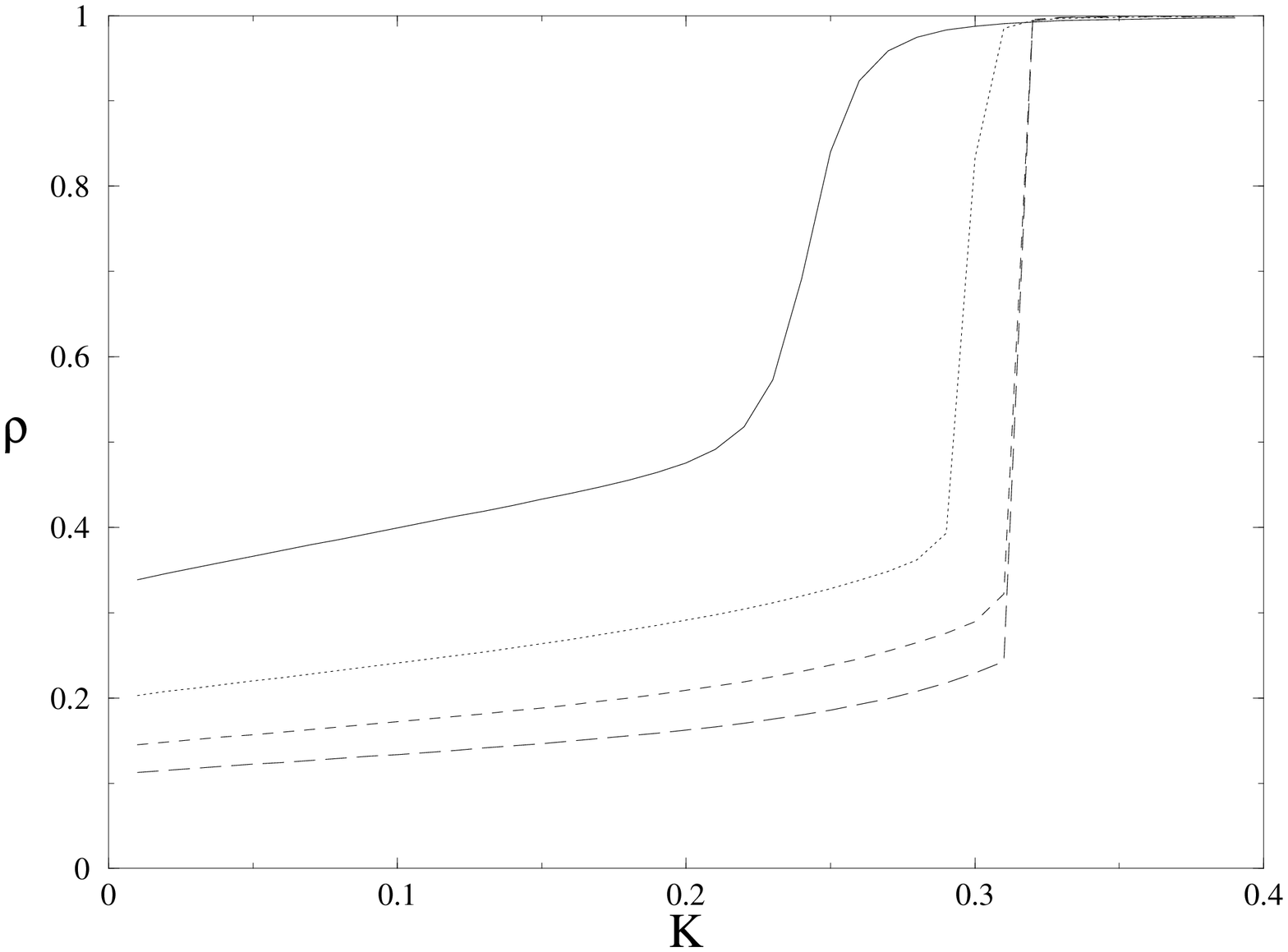}
\end{center}
\caption{Plot of density $\rho$ for 
$\beta=1.1$} \label{denspl1b}
\end{figure}

Let us now characterise the different phases and transitions.
For low
$K$ the average length of polymer is finite, the density of the
polymer on the lattice must therefore be zero. 
The first transition as $K$ is increased is to one of two dense
phases, depending on whether $\beta<\beta_H$ or $\beta>\beta_H$. It is
convenient to use the density as an order parameter. For practical
reasons we choose boundary conditions in which the polymer is 
made to extend the length
of the lattice, between
$x=0$ and $x\to \infty$. 
This obliges the polymer to be infinite in length even in the
zero-density phase. 
A choice of boundary conditions is not expected to affect
the critical behaviour of an equilibrium system. 
The partition function becomes ${\cal
Z}=\lim_{x\to\infty} \lambda_1^x$ for both the high and low density
phases, and $\rho$ may be calculated using \eref{normrho} in all
phases. 
The
density will now be finite for $K<K_c$ as long as the
lattice width is finite, but
must tend to zero as $L\to\infty$. This is indeed what is observed
from the density plots shown in figures~\ref{denspl1a} and~\ref{denspl1b}.

\begin{figure}
\begin{center}
\includegraphics[width=7.5cm,angle=-90]{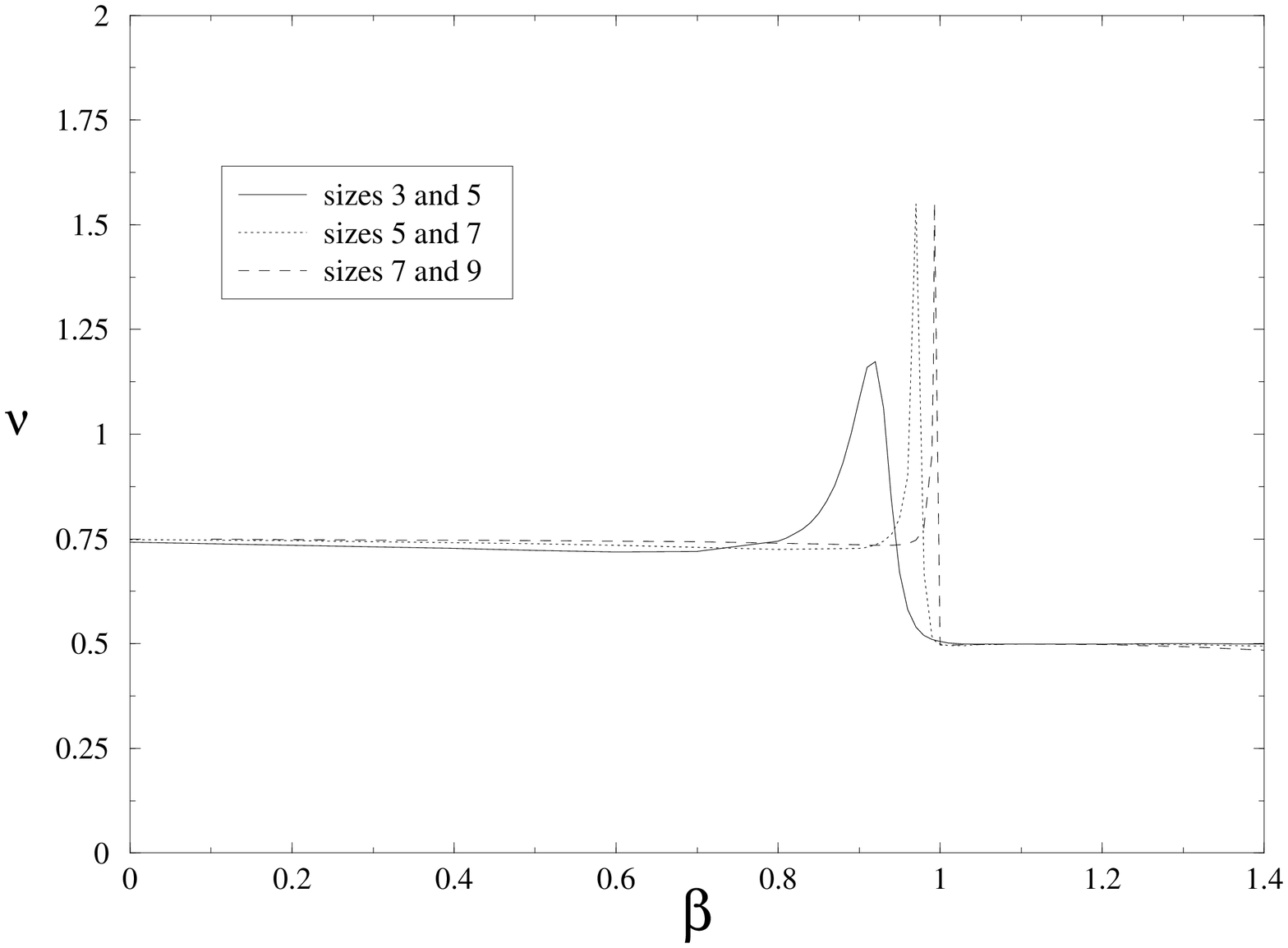}
\caption{Estimates for the exponent $\nu$ for the low-$K$ transition line}
\label{oddnu}
\end{center}
\end{figure}

\begin{figure}
\begin{center}
\includegraphics[width=6.5cm,angle=-90]{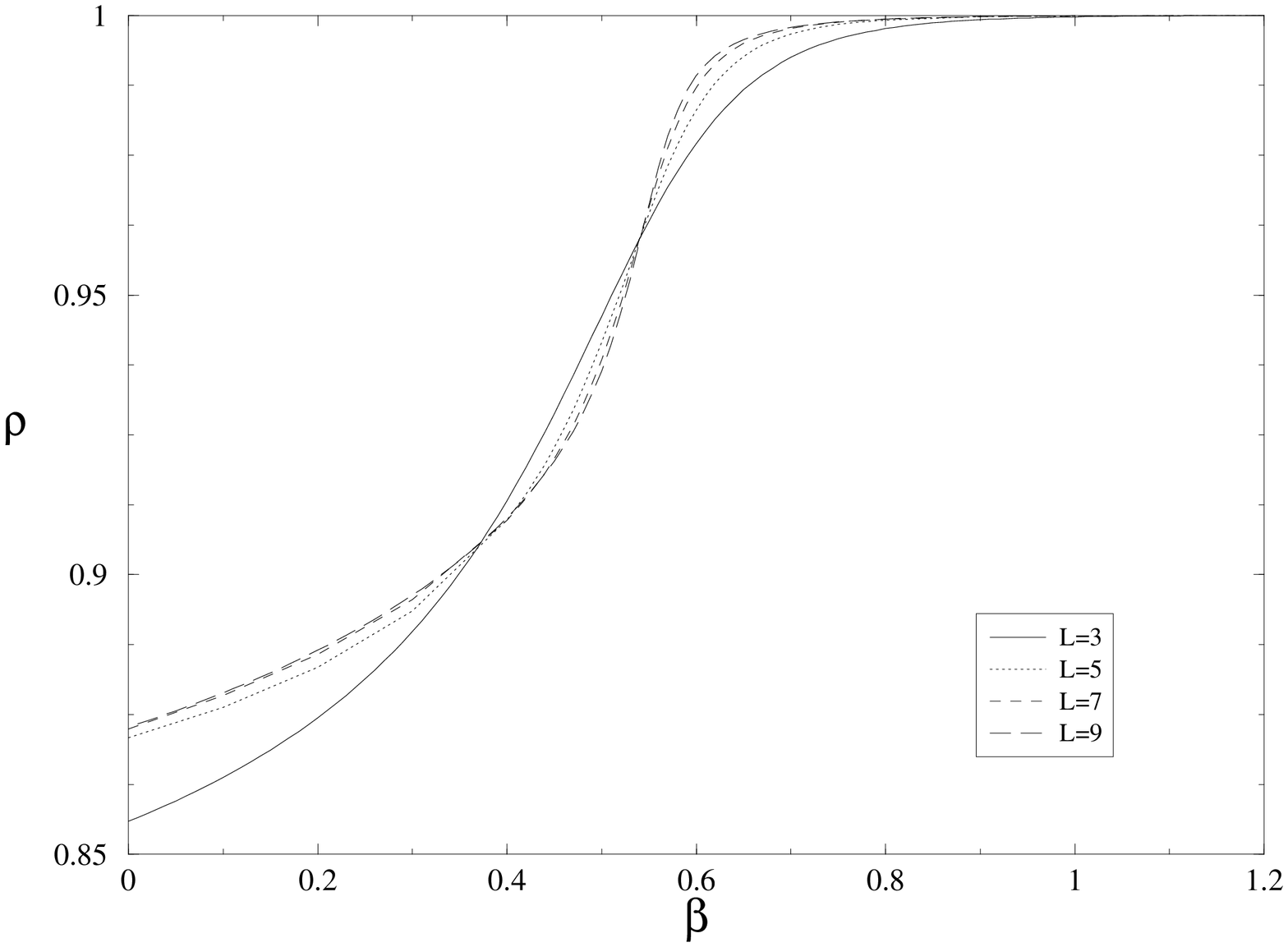}\\
\includegraphics[width=6.5cm,angle=-90]{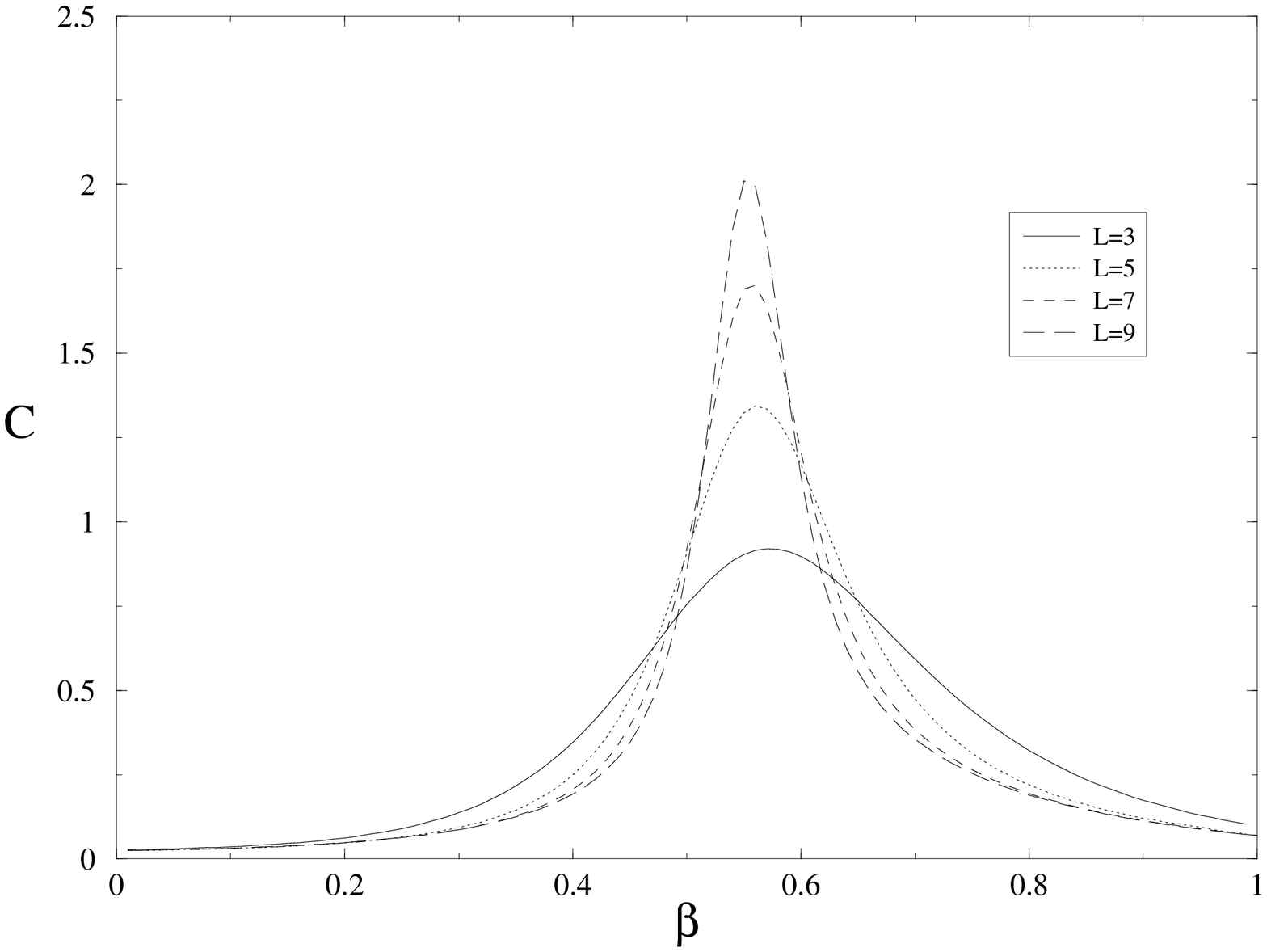}\\
\includegraphics[width=11.cm]{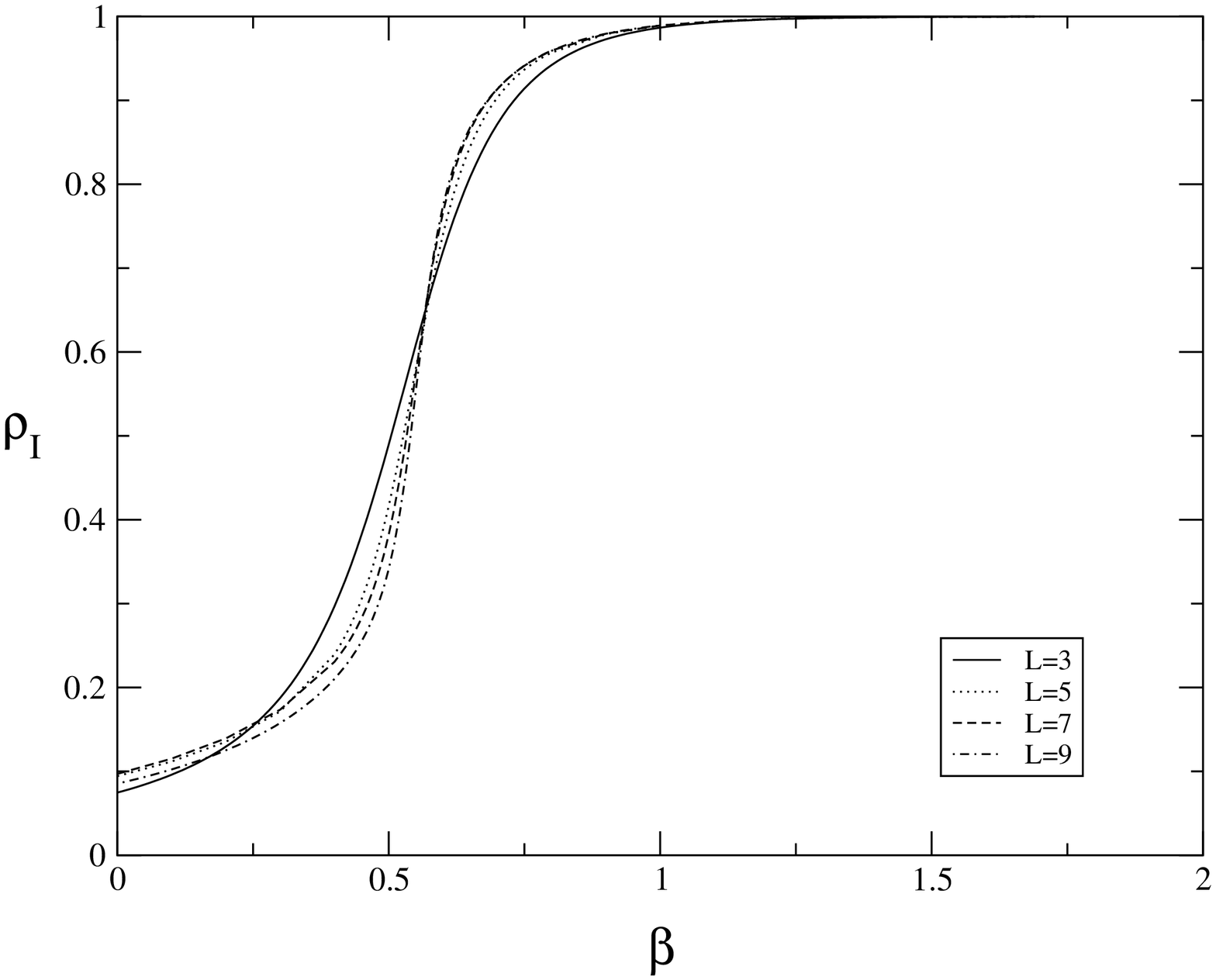}
\end{center}

\caption{Plots of the density $\rho$ (top), the
specific heat $C$ (middle) and the density of interactions $\rho_I$ 
(bottom) for
$K=1.5$}\label{denspl2}
\end{figure}

For $\beta<\beta_H$ the behaviour of $\rho$ and the length flucuations
indicate the existence of a line of second order 
phase transitions, as expected.
The line 
is in the  self-avoiding walk universality
class, having $\nu=3/4$ as may be seen from 
the finite
size estimates for $\nu$ shown in \fref{oddnu}. 
The
transition is to a finite density phase. 

For
$\beta>\beta_H$ the density  plot, given in
\fref{denspl1b},
indicates that the
transition is first order  to a dense phase. 
That this line corresponds to the
compact polymer state is confirmed by the observed
exponent $\nu=0.5=1/d$, see \fref{oddnu}. 
A major difference with the $\theta$-point model,
seen using
the phenomenological renormalisation scheme is the existence of
two different high density phases. The density plots in
\fref{denspl1b} 
for
$\beta>\beta_H$ suggest that the walk changes discontinuously from
$\rho=0$ to $\rho=1$, and that the density
of interactions 
saturates instantly. 
In other words the lattice changes abruptly from
being essentially 
empty to essentially full, with a maximum density of interactions. 
Trivially, a lattice which is empty (a finite walk length on an
infinite lattice)
has zero energy and zero entropy per site; 
the free energy per site is zero. 
If the lattice is completely
full with the interactions density saturated, the walk may only be in
one of two configurations; 
either all the bonds are vertical or all the
bonds are horizontal. The entropy per site in
this phase is therefore also zero.
Since, in this state, there is one bond and one interaction
per site, the energy per site is $e=\mu-\epsilon$. The transition
between the two phases would then occur when the free energy per site,
$f=e-Ts=e=0$, in other words when $\mu=\epsilon$. 
Remembering the 
choice $\epsilon=1$, the transition line would be 
$K=\exp(-\beta)$. This line has been plotted 
for comparison on the phase
diagrams, shown in \fref{prgpd}.
The agreement with the observed
first-order transition line is excellent.

In order to differentiate between the two high density phases, it is
convenient to use the interaction density, $\rho_I$,
  as an order
parameter. 
This order parameter,
along with its fluctuations, is plotted as a function of
$\beta$ for
$K=1.5$ in \fref{denspl2}. The order parameter differentiates between
the two phases since it is identically one in the high-$\beta$ phase,
and less than one 
in the low-$\beta$ phase. The variation of the order
parameter appears to be fairly smooth for all lattice sizes, and leads
one to conjecture that this high density transition is critical. 
Also
plotted in \fref{denspl2} is the specific heat, calculated as the
second derivative of the free energy with respect to the temperature,
keeping $\mu$ and $\epsilon$ constant.

Various estimates of the critical $\beta$ for $K=1.5$, calculated using the
phenomenological renormalisation group and by locating the peaks in
the specific heat, are given in \tref{horiztabl}.  The various results
are coherent with an estimate $\beta_c(K=1.5)=.55\pm 0.01$.
At a critical transition, the
specific heat diverges as $T\to T_c$ with an exponent $\alpha$,
\begin{equation}
C\sim|T-T_c|^{\alpha}.
\end{equation}
When combined with relation \eref{corsc} and $\xi_L\propto L$, this 
leads to the finite-size scaling result
\begin{equation}
C_{\rm max}\approx
A+B L^{\alpha/\nu}.
\end{equation}
A three point fit for odd widths gives $\alpha/\nu=0.67$ for $L=3,5,7$
and $\alpha/\nu=0.55$ for $L=5,7,9$. A similar three point fit for the
even sizes, $L=4,6,8$ gives a value of $\alpha/\nu=0.156$. If
sufficiently large lattice widths are considered then $A$ can be
dropped. When this is not the case, then the result is an effective
value of the exponent, which should converge to the correct value as
the lattice width is increased. Unfortunately the number of two point
approximations we could construct does not allow for meaningful
extrapolation. All we could determine is that $0.156\leq
\alpha/\nu\leq 0.55$ gives a reasonable limit on the possible values
of $\alpha/\nu$. Using the scaling relation $\alpha=2-d\nu$, this gives
$\nu$ as $0.78\leq \nu \leq 0.93$. Note that this range is coherent with the
values of $\nu$ found from the phenomenological renormalisation group
calculation, shown in \tref{horiztabl}. Additionally, the upper
boundary, calculated from the even lattice sizes, is in good agreement
with the even lattice size estimates of $\nu$ in \tref{horiztabl}.

\begin{table}
\caption{Estimates for (a) $\beta_c$ and $\nu$ for $K=1.5$
limit
using phenomenological renormalisation group with strips of width $L$
and $L+2$ and (b) $\beta_c$ as estimated from the peak of the specific
heat and the height of the peak}
\label{horiztabl} 
\begin{indented}
\item\begin{tabular}{@{}lllll}
\br
 & \centre{2}{\bf(a)} & \centre{2}{\bf (b)} \\
 & \crule{2} & \crule{2}\\
$L$ & $\beta_c$ & $\nu$ 
& $\beta^L_c$ & $C_{\rm max}$\\
\mr
3 & 0.579774 & 0.885482 &  0.572447 & 0.920518 \\
4 & 0.546148 & 0.920203 & 0.500045 & 0.645603 \\
5 & 0.560960 & 0.841050  & 0.561950 & 1.344506 \\
6 & 0.550013&0.920194 &  0.510141 & 1.131212  \\
7 & 0.554531 & 0.864167& 0.556259 &  1.706669\\
8 & -- & -- &  0.524612 & 1.494860 \\
9 & -- & -- &  0.553865 & 2.024505 \\ 
\br
\end{tabular}
\end{indented}
\end{table}

\begin{figure}
\begin{center}
\includegraphics[width=7.5cm,angle=-90]{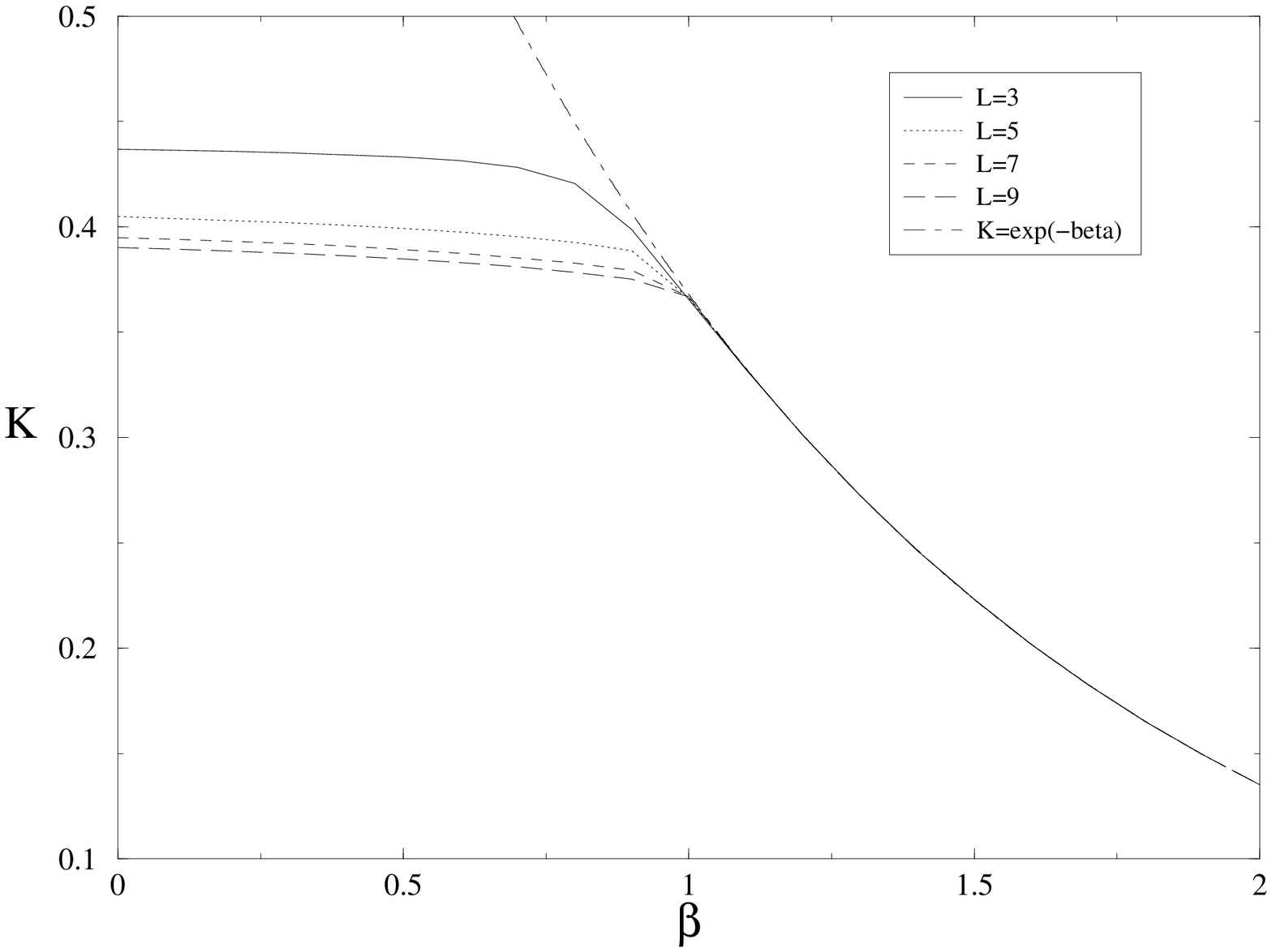}\\
\includegraphics[width=7.5cm,angle=-90]{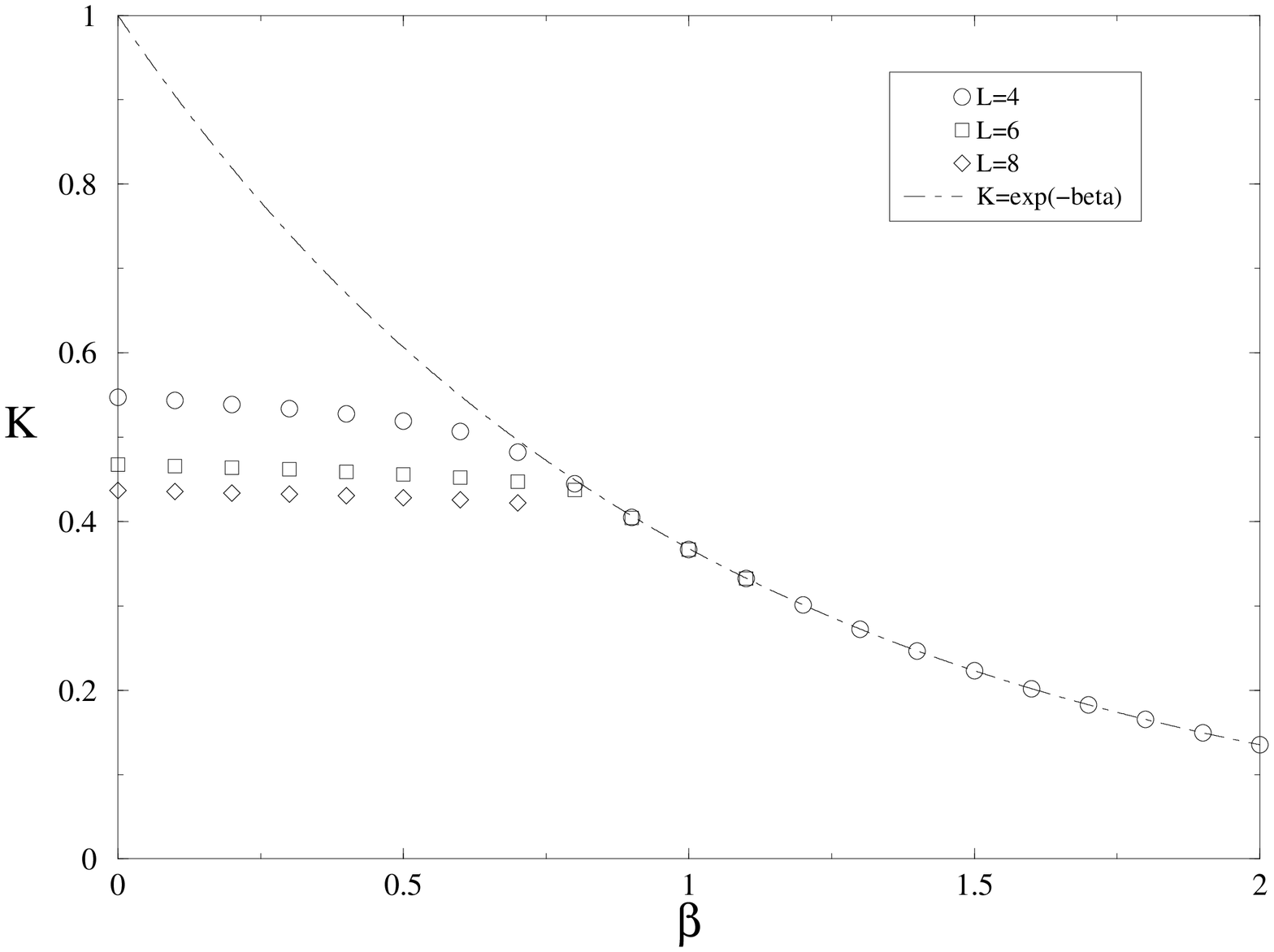}
\end{center}
\caption{Phase diagram estimates calculated using the condition
that the largest eignvalue $\lambda_1=1$ odd lattice widths (top) and
even lattice widths (bottom)}\label{lampd} 
\end{figure}

\begin{figure}
\begin{center}
\includegraphics[width=7.5cm,angle=-90]{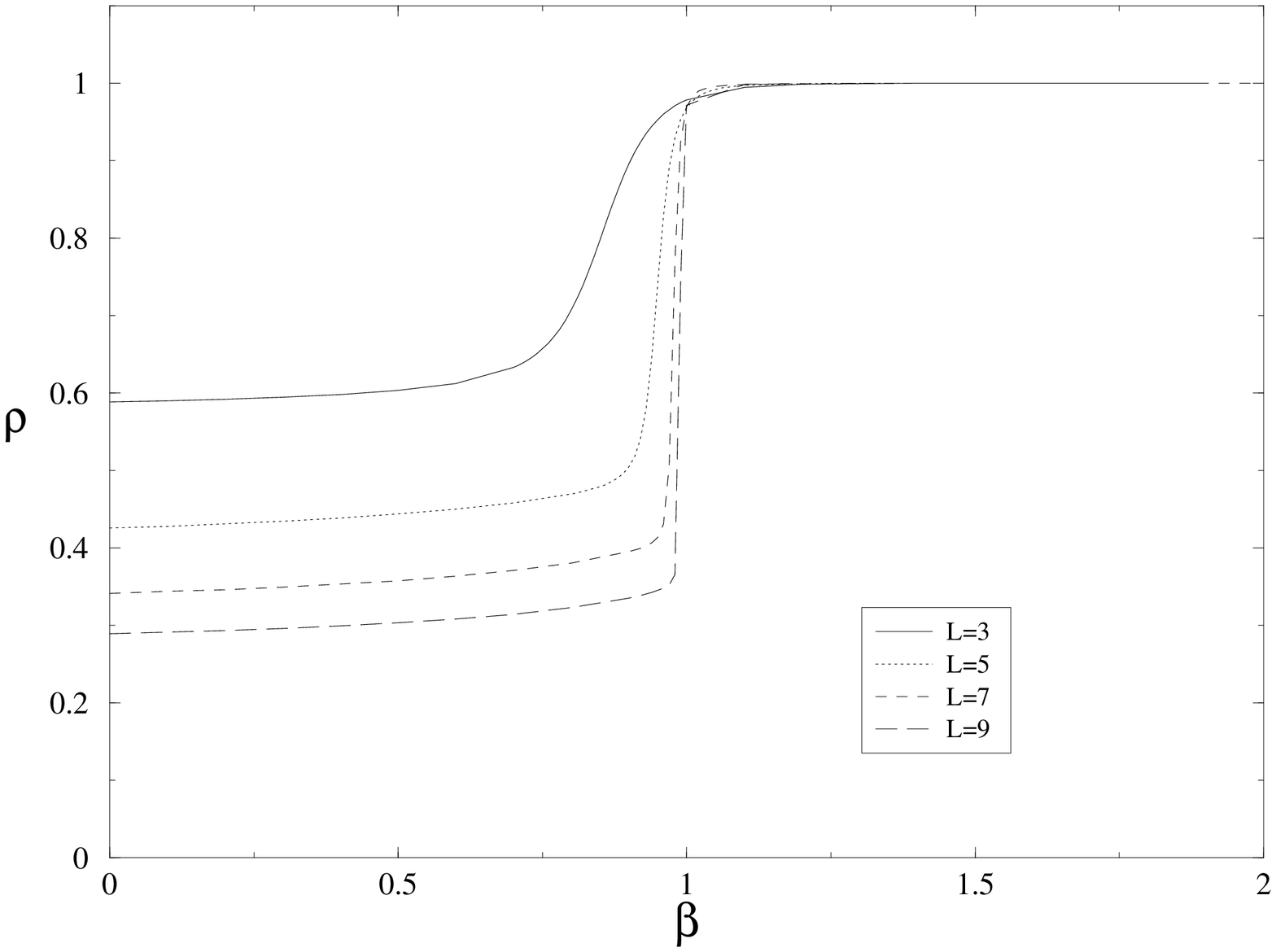}
\includegraphics[width=7.5cm,angle=-90]{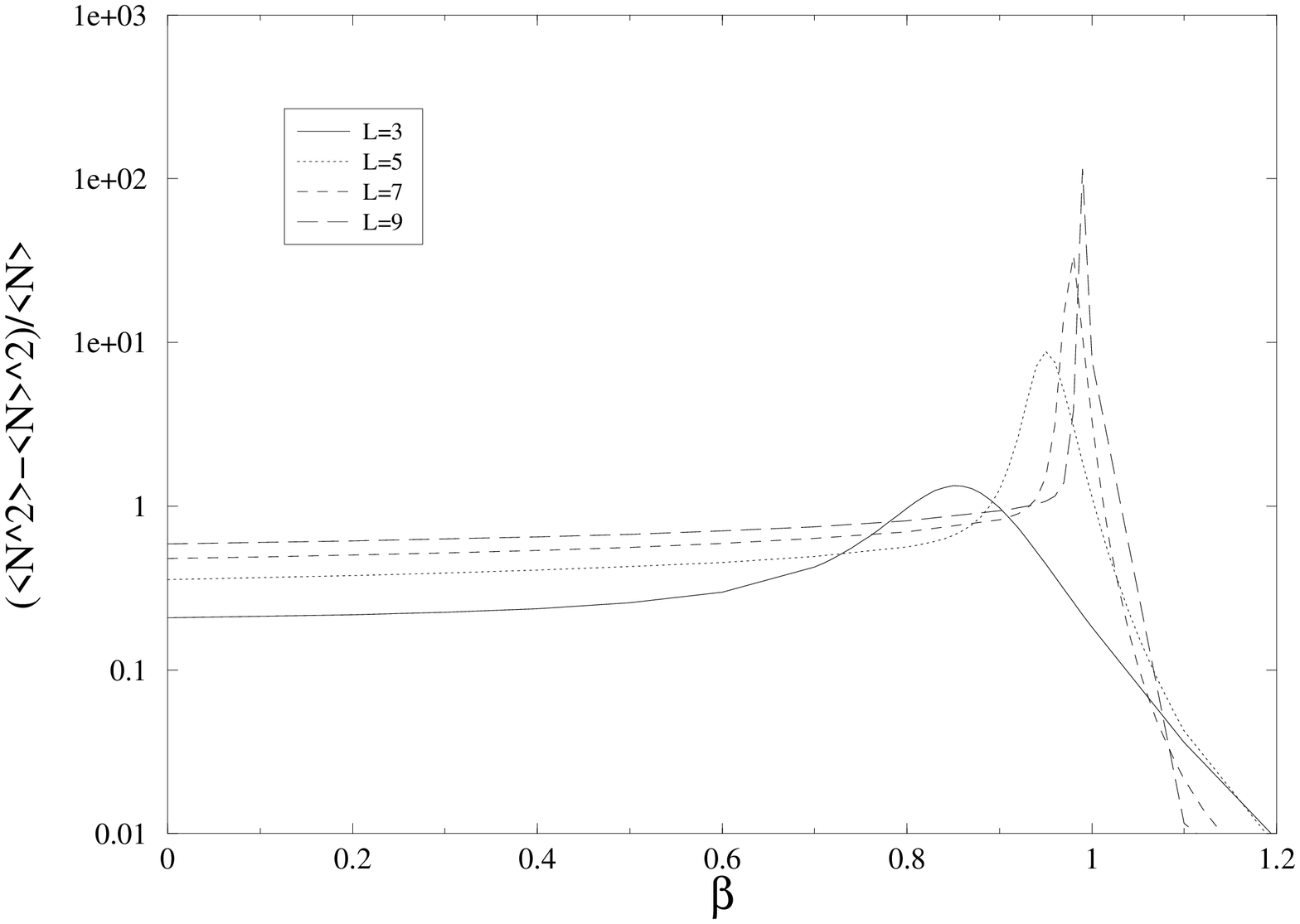}
\end{center}
\caption{Density (top) and density 
fluctuations (bottom) along the transition line as estimated
using the condition that the largest eigenvalue $\lambda_1=1$}\label{lamdens}  
\end{figure}


It now remains to determine the nature of the transition at
$(K_H,\beta_H)$, where the three transition lines join. For the
standard $\theta$ point model, this point is tricritical, and
corresponds to the confluent of two transition lines, not three. 

\begin{figure}
\begin{center}
\includegraphics[width=7.5cm,angle=-90]{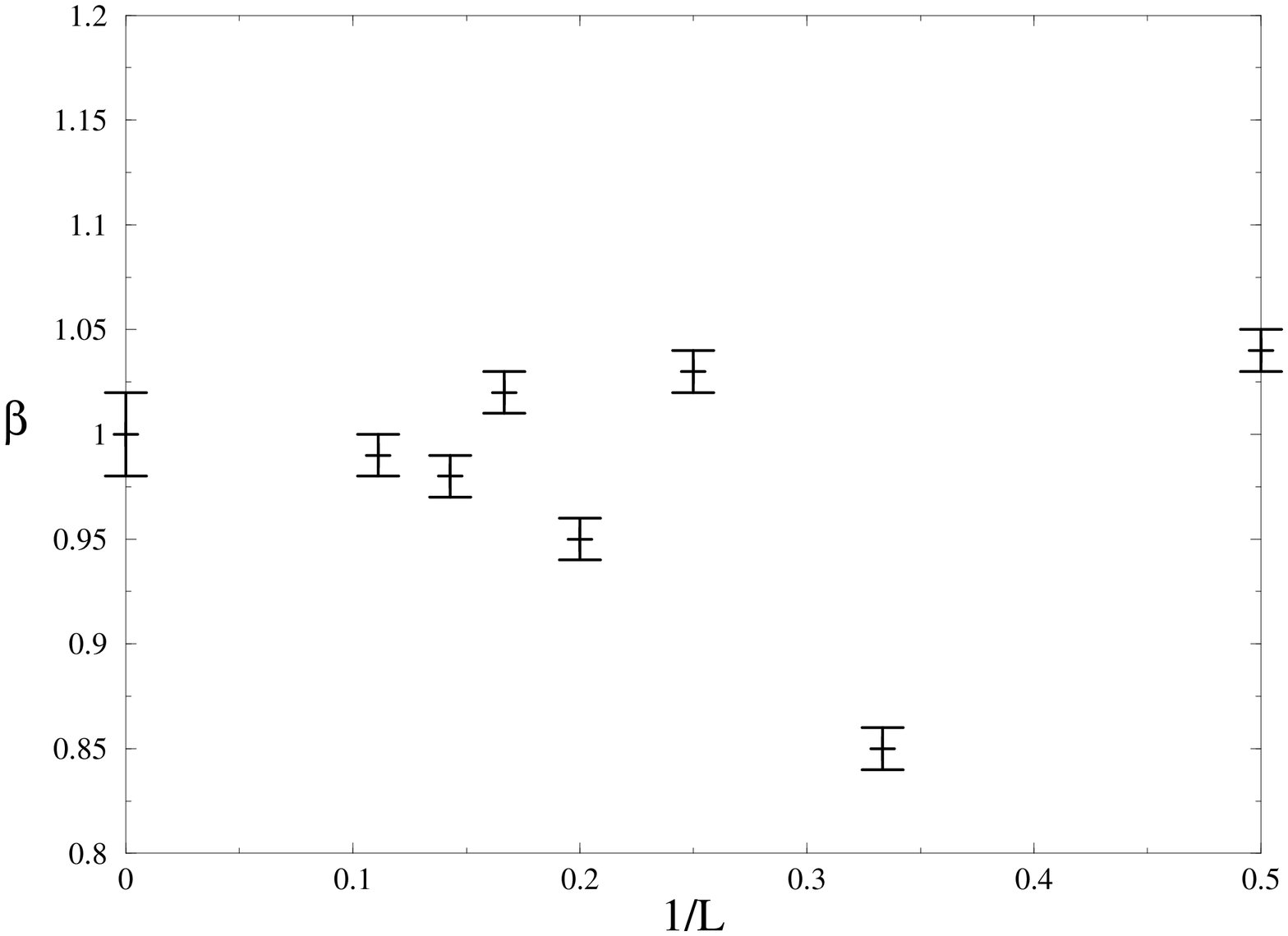}
\end{center}
\caption{Finite size estimates of $\beta_H$, the value of $\beta$ at
the collapse transition. The upper points, calculated for even lattice
widths, come from the estimated position at which the two critical
lines merge. The lower points, calculated for the odd lattice sizes,
come from the peaks of the fluctuations in $N$. The error bars given
correspond to the interval between the points calculated. 
These error bars could be reduced at an additional cost in computer
time.}\label{bHest}
\end{figure}

The problem we are faced with is that the phenomenological 
renormalisation group applied to even lattice widths does not give the
first-order transition at all. When applied to the odd lattice widths,
the method is not able to follow the self-avoiding walk transition
line all the way to $\beta_H$, jumping to the upper transition line
just short of $\beta_H$. This undershoot rapidly becomes smaller as
the lattice width is increased. 
Remembering that the length of the walk first diverges when
$\lambda_1\to 1$ and that this condition coincides with the low-K
phase transitions in the thermodynamic limit, we can use the
condition $\lambda_1(K,\beta)=1$ as a criterion for defining finite
width estimates to the transition lines. The phase diagram estimates 
calculated
using this criterion are shown in \fref{lampd}.
This method is incapable of
giving the high-$K$ transitions, and converges more slowly 
than the
phenomenological renormalisation group,
but has the advantage that the transition lines are fairly smooth and
continuous. 
The density and its fluctuations, calculated along
this line, are presented in
\fref{lamdens}. As the width of the lattice is increased, the jump in
the density becomes sharper. The height of the fluctuations (shown
with a log-linear scalin in \fref{bHest}) increases extreemly rapidly
with lattice width, stronly suggesting a delta peak in the
thermodynamic limit. Whilst the range of lattice widths prevent us
from being categorical, the results suggest strongly 
that the transition is
first order along the transition line. This seems to confirmed in the
phase diagram. In both figures~\ref{prgpd} and~\ref{lampd} the
transition line has a discontinuous slope at $\beta_H$.
  
The positions of the peaks of density fluctuations
are shown in \fref{bHest} along with the positions of the cusps where the
upper and lower critical lines join for even lattice widths. 
We estimate
$\beta_H=1.00\pm 0.02$, with $K_H=\exp(-\beta_H)$.
It is curious that $\beta_H$ should be so close to one, though
we have as yet not been able to 
determine why
this should be the case.

\section{The Hamiltonian Walk limit ($K\to\infty$)}\label{hamilt}

In the limit $K\to\infty$ the walk will fill the lattice
maximally. Each site will be visited  once and only once. 
In this limit the walk becomes what is known as an Hamiltonian 
Walk\cite{Barber}. 
In this limit the 
effect of the interaction is to energetically penalise corners in the
walk. This is similar to the Flory model\cite{hs} except 
that there 
are configurations which introduce one or two corners into the walk
with no difference in energy, see \fref{interact}.

\begin{figure}
\begin{center}
\includegraphics[width=7.5cm,angle=-90]{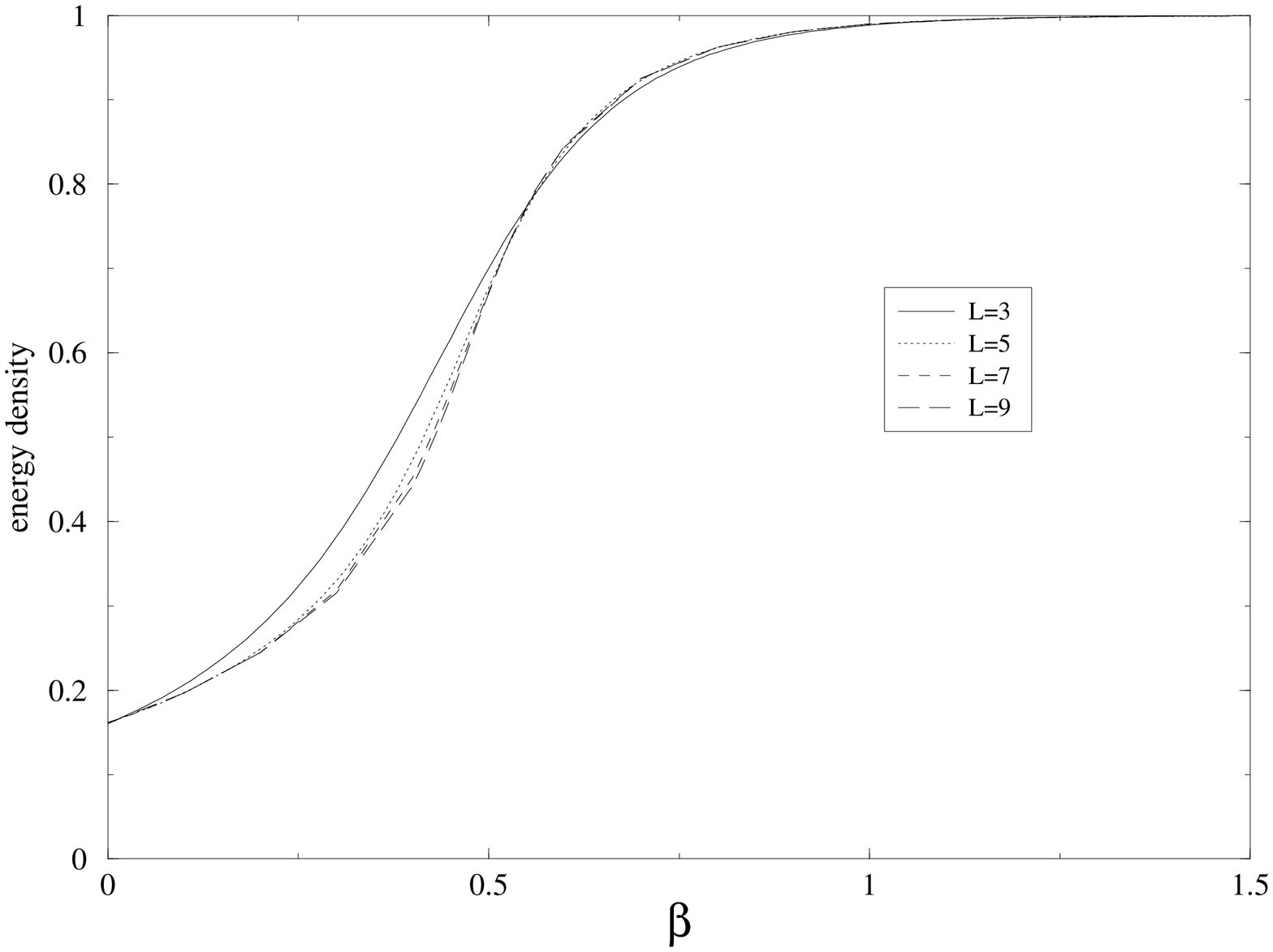}\\
\includegraphics[width=7.5cm,angle=-90]{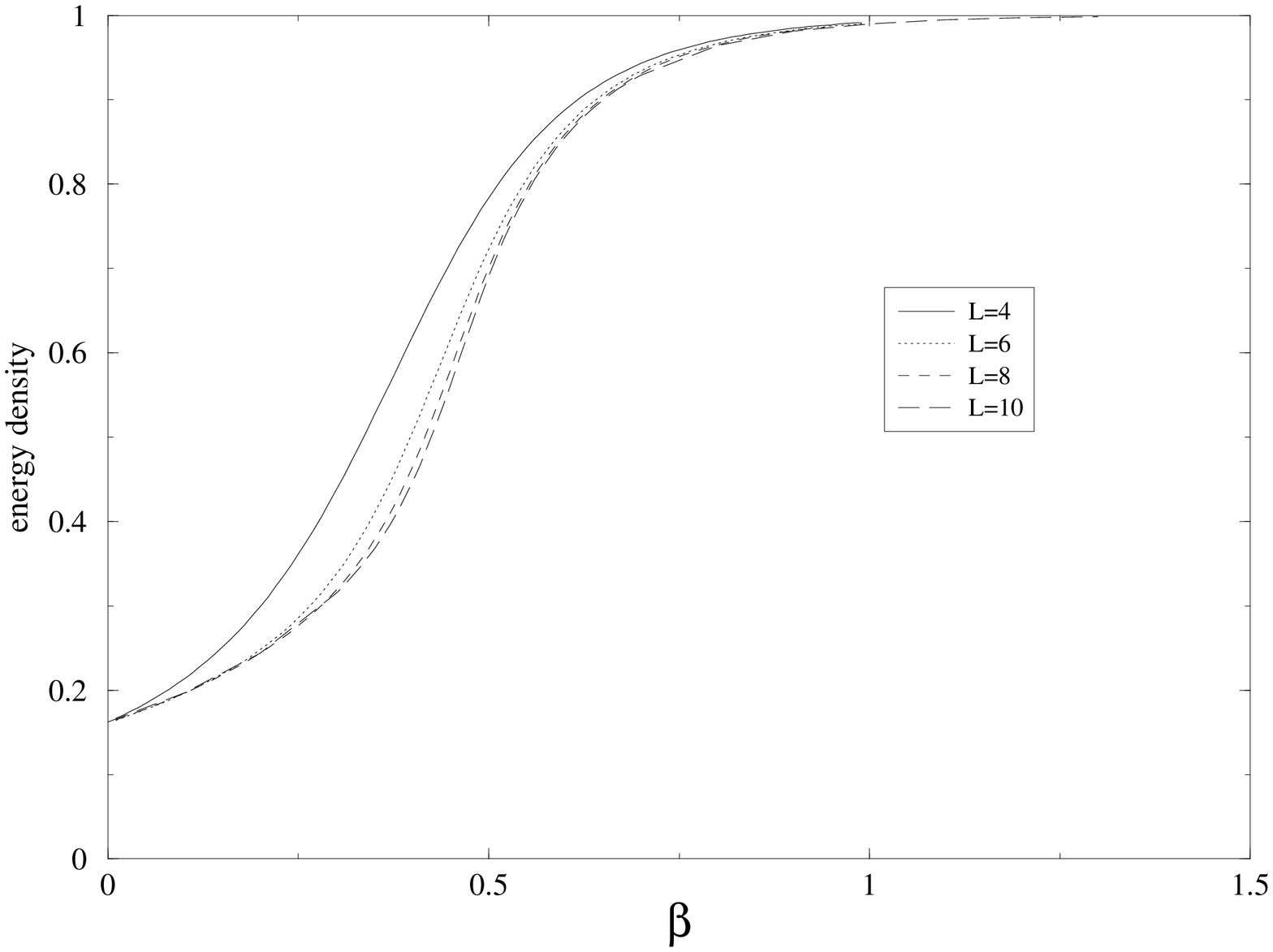}
\end{center}
\caption{Energy
density
as a function of $\beta$ in the
$K\to\infty$ limit for odd lattice widths (top) and even lattice
widths (bottom)}\label{corndens}
\end{figure}

There is a low-temperature corner-free phase. As the temperature is
raised there is a transition in which the density of corners becomes
non-zero. To distinguish between the high and low temperature phases
it is possible to use either the density of corners 
the energy density.
This latter  is shown, as a function of $\beta$
in \fref{corndens}.

\begin{table}
\caption{Estimates for (a) $\beta_c$, $\nu$ and $\eta$ in the $K\to\infty$
limit
using phenomenological renormalisation group with strips of width $L$
and $L+2$ and (b) $\beta^L_c$, the critical value of $\beta$ 
as estimated from the peak of the
interaction fluctuations, the height of the peak ($f_{\rm max}$) and
the value of $\eta$ calculated the extrapolated value of $\beta_c$. 
Even and odd lattice widths are taken
apart}
\label{denstrans} 
\begin{indented}
\item\begin{tabular}{@{}lllllll}
\br
 & \centre{3}{\bf(a)} & \centre{3}{\bf (b)} \\
 & \crule{3} & \crule{3}\\
$L$ & $\beta_c$ & $\nu$ & $\eta$ 
& $\beta^L_c$ & $C_{\rm max}$& $\eta(\beta_c^\infty)$\\
\mr
3 & 0.450472 & 1.109084 & 0.082484
& 0.435221 &  1.714121 & 0.087226
\\
4 & -- & -- & -- 
&  0.375742 & 1.875099 & 0.035357 \\
5 & 0.461359 &  1.060628 & 0.085672 
& 0.465233 & 2.124762 & 0.090245\\
6 & -- & -- & -- 
& 0.437435 & 2.260666 & 0.047824\\
7 & 0.465970 & 1.105916 & 0.087611 
& 0.471160 & 2.354970 & 0.092053\\
8 & -- & -- & -- & 0.454834 & 2.491249 & 0.056830  \\
9 & -- & --  & -- & 0.473274 & 2.492086&0.093240 \\
10 & -- & -- & -- & 0.461866 & 2.636003 & 0.064064\\
\mr 
$L\to\infty$ & 0.477 & -- & -- 
& $0.476\pm 0.001$ & 2.867 & $0.096\pm0.002$\\
(odd)& & & & & &\\
$L\to\infty$  & -- & -- & -- 
& $0.473\pm 0.003$ & 3.071 & $0.10\pm0.01$ 
\\
(even)& & & & & &\\
\br
\end{tabular}
\end{indented}
\end{table}

\begin{figure}
\begin{center}
\includegraphics[width=7.5cm,angle=-90]{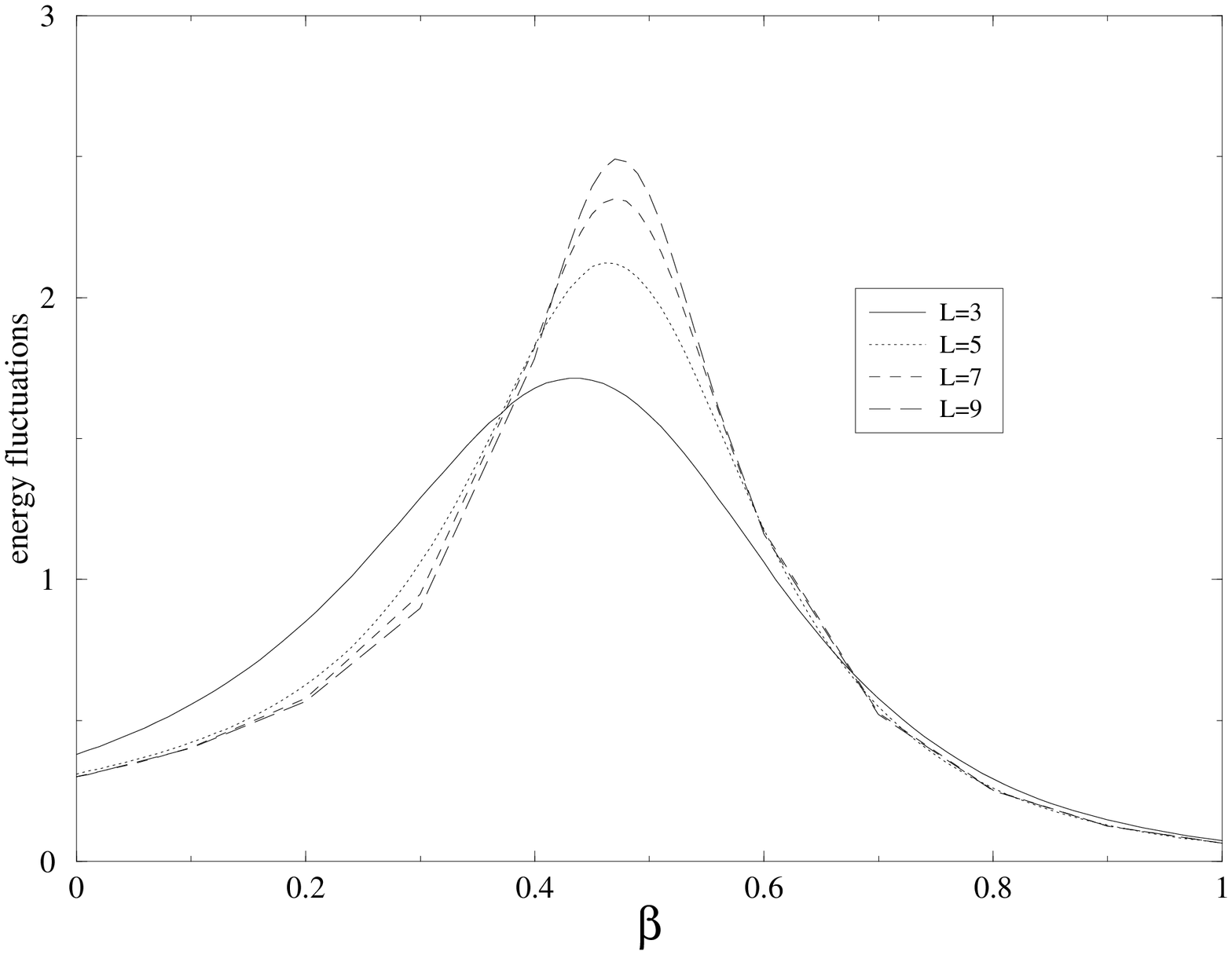}\\
\includegraphics[width=7.5cm,angle=-90]{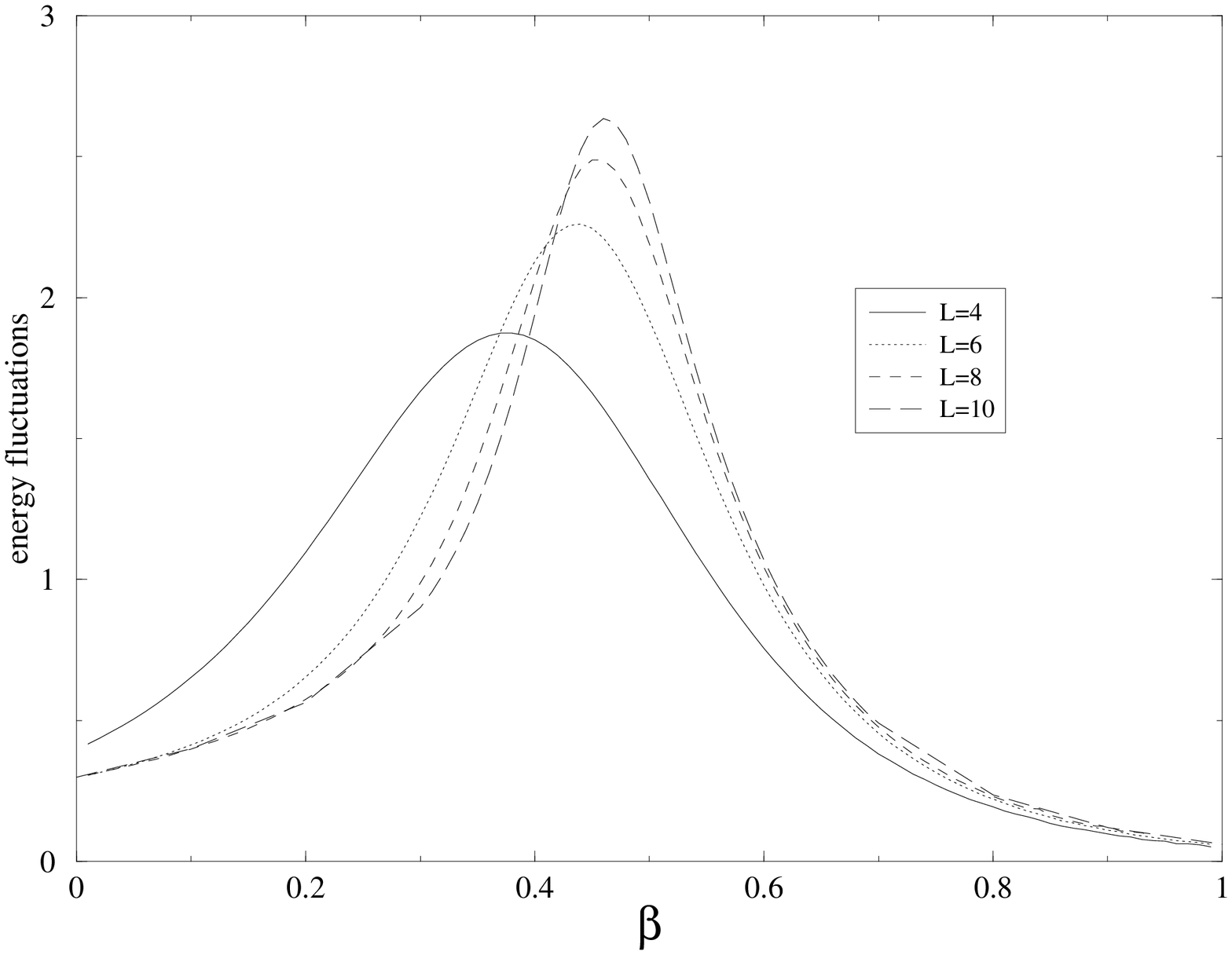}
\end{center}
\caption{The energy fluctuations in the $K\to \infty$
limit for odd lattice widths (top) and even lattice
widths (bottom)}\label{densfluct} 
\end{figure}

We estimate the critical temperature again using the 
phenomenological
renormalisation group, the results of which are shown in
\tref{denstrans} and extrapolated using a three point fit to
\begin{equation}
\beta^{\infty}_c=\beta_c^L+A L^B,
\end{equation}
where $\beta_c$ is the finite size estimate and $\beta_c^{\infty}$ is
the extrapolated value. $A$ and $B$ are constants to be determined.

Only the odd lattice widths presented solutions to
the phenomenological renormalisation group equation \eref{fss}, 
given in \tref{denstrans}.
The positions and values of the maxima of the energy fluctuations 
per site, 
shown in \fref{densfluct}, are also reported in
\tref{denstrans}. The fluctuations appear to 
saturate. We assume that, even if the fluctuations do not
diverge, the transition point is estimated by the value of $\beta$
for which the fluctuations are maximal. This agrees well with the
phenomenological renormalisation group estimate, and fits with a 
continuation of the finite $K$ phase diagrams (\fref{prgpd}) 
to high values of $K$.
Phenomenological renormalisation and the peaks of the energy
fluctuations give 
estimates of 
the critical temperature which concur, giving
 $\beta_c=0.476\pm 0.001$. 

It is also possible to estimate the exponant $\eta$, using 
the conformal invariance result for periodic
boundary conditions\cite{cardy}:
 \[
\eta=\frac{L}{\pi\xi}.
\]
We find
$\eta_c=0.096\pm0.002$, see \tref{denstrans}. 

It is anticipated that the entire high
temperature phase is critical, in analogy with other models of this
type, such as the model F\cite{modf}, the Flory model\cite{hs} 
and the loop gas model on
the brickwork lattice\cite{egfos}. 
The
estimates of $\nu$ are not at all clear. They seem  close to 
$\nu=1$, but 
in other respects the transition is similar to that
seen in the model F
and the Flory model, which have transitions of infinite order with $\nu$
infinite.

\section{Conclusions}\label{CONC}

In this article we have presented a two-dimensional model for
homopolymer collapse under the influence of ``hydrogen-bonding'' like 
nearest-neighbour interactions. The phase diagram is quite 
different from the equivalent $\theta$ point phase diagram; the
collapse transition  is now first order. This is also observed in the
oriented interacting self-avoiding walk model, where, for appropriate values of
the model parameters, a first order
collapse occurs to a spiral configuration\cite{trovseno}. 
These two models have in
common that the collapsed phase is anisotropic, in our case one of the
two lattice directions is selected, and in the oriented walk case, for
strong enough parallel interaction strength, one of two chiralities is
selected. In the standard $\theta$ model the collapsed phase is
isotropic.

Here we also observe an additional
transition in the dense walk regime. This transition appears to be
second order, though for the moment we have not been able to determine
to any accuracy any of the critical exponents. Similar critical lines
are observed in the loop models with corner
interactions, with or without a collapse transition\cite{egfos,nien}. 

There also exists non-trivial critical behaviour in the
$K\to\infty$ limit, where there is a finite temperature phase
transition between a critical high temperature phase and a frozen low
temperature phase. 

Another well studied model with a qualitatively similar phase diagram 
is the partially directed interacting polymer
model\cite{directed,wt1}. There are major differences however: The
line $K=\exp(-\beta)$ corresponds to a phase transition line for all
values of $K$, deliminating a similar fully collapsed phase. 
The equivalent phase to the finite-$\rho$ phase
has a density $\rho=0$ due to the directed nature of the model. The
equivelent to the theta point  transition is second
order\cite{directed} (here it appears to be first order).

\pagebreak

\ack

DPF would like to thank the kind hospitality of the INFM and the
Dipartimento di
Fisica, Universit\`a di Padova during the early stages of this work.

\end{document}